\documentclass[sigconf]{acmart}
\AtBeginDocument{%
  \providecommand\BibTeX{{%
    \normalfont B\kern-0.5em{\scshape i\kern-0.25em b}\kern-0.8em\TeX}}}

\copyrightyear{2021}
\acmYear{2021}
\setcopyright{iw3c2w3}
\acmConference[WWW '21 Companion]{Companion Proceedings of the Web Conference 2021}{April 19--23, 2021}{Ljubljana, Slovenia}
\acmBooktitle{Companion Proceedings of the Web Conference 2021 (WWW '21 Companion), April 19--23, 2021, Ljubljana, Slovenia}
\acmPrice{}
\acmDOI{10.1145/3442442.3452347}
\acmISBN{978-1-4503-8313-4/21/04}

\begin{document}

\title{Language-agnostic Topic Classification for Wikipedia}

\author{Isaac Johnson}
\email{isaac@wikimedia.org}
\affiliation{%
  \institution{Wikimedia Foundation}
  \country{United States}
}

\author{Martin Gerlach}
\email{mgerlach@wikimedia.org}
\affiliation{%
  \institution{Wikimedia Foundation}
  \country{United States}
}

\author{Diego S\'{a}ez-Trumper}
\email{diego@wikimedia.org}
\affiliation{%
  \institution{Wikimedia Foundation}
  \country{United States}
}

\renewcommand{\shortauthors}{Johnson et al.}

\begin{abstract}
  A major challenge for many analyses of Wikipedia dynamics---e.g., imbalances in content quality, geographic differences in what content is popular, what types of articles attract more editor discussion---is grouping the very diverse range of Wikipedia articles into coherent, consistent topics. This problem has been addressed using various approaches based on Wikipedia's category network, WikiProjects, and external taxonomies. However, these approaches have always been limited in their coverage: typically, only a small subset of articles can be classified, or the method cannot be applied across (the more than 300) languages on Wikipedia. In this paper, we propose a language-agnostic approach based on the links in an article for classifying articles into a taxonomy of topics that can be easily applied to (almost) any language and article on Wikipedia. We show that it matches the performance of a language-dependent approach while being simpler and having much greater coverage.
\end{abstract}



\maketitle

\section{Introduction}
As of January 2021, Wikipedia has over 300 language editions with 55.7 million articles\footnote{\url{https://wikistats.wmcloud.org/display.php?t=wp}} about 20.4 million distinct entities\footnote{Personal calculation based on 4 January 2021 Wikidata JSON dump: \url{https://dumps.wikimedia.org/wikidatawiki/entities/20210104/}} and an additional 250 thousand articles created every month.\footnote{\url{https://stats.wikimedia.org/\#/all-wikipedia-projects/contributing/new-pages/normal|bar|2-year|~total|monthly}} These articles cover a very wide range of content and it can be difficult to track and understand these dynamics---e.g., what types of content are attracting the interest of editors or readers?---especially across the different language editions.

Wikipedia itself has a number of editor-curated annotation systems that bring some order to all of this content. Most directly perhaps is the category network, but content can also be categorized based on properties stored in Wikidata, tagging by WikiProjects (groups of editors who focus on a specific topic), or inclusion of templates such as infoboxes. While these annotation systems are quite powerful and extensive, they ultimately are human-generated and semi-structured and thus have many edge cases and under-coverage in languages or communities that do not have editors who can maintain these annotations (see~\cite{hall2017freedom}).
Researchers have developed many approaches to improve these annotation systems by using Wikipedia's category network~\cite{piccardi2018structuring,lewoniewski2019multilingual}, WikiProjects~\cite{asthana2018few,warncke2015misalignment}, depending on DBPedia's manually-curated taxonomy that then assigns topics based on infobox templates~\cite{lewoniewski2019multilingual}, or throwing out the editor-based annotation systems completely and learning a set number of topics through unsupervised techniques such as topic modeling~\cite{mimno2009polylingual,piccardi2020crosslingual,singer2017we,miz2020trending}.

However, these approaches generally suffer from two limitations related to coverage in terms of language and article. 
First, not all language communities have the editor base or need to maintain these annotations to the same degree. For example, Arabic Wikipedia has the most categories per article at 25.5, and English Wikipedia with its approximately 40,000 monthly active editors,\footnote{\url{https://stats.wikimedia.org/\#/en.wikipedia.org/contributing/active-editors/normal|line|2-year|(page_type)~content*non-content|monthly}} has 1.5M categories that are collectively applied 66M times across its 6.2 million articles.\footnote{Personal calculations using the December 2020 dumps and categorylinks table (\url{https://www.mediawiki.org/wiki/Manual:Categorylinks_table}) filtered by page table to articles in namespace 0 (\url{https://www.mediawiki.org/wiki/Manual:Page_table)}} Wu Chinese Wikipedia, however, with approximately 20 active editors and 41,231 articles only has 6,990 categories that are applied 26,396 times (leaving many articles with no categories at all). Similar variation is seen with other systems as well: linkage to Wikidata is higher but still as low as 87.5\% in Cebuano Wikipedia with its 5.5M articles\footnote{\url{https://wikidata-analytics.wmcloud.org/app/WD_percentUsageDashboard}}, only 92 Wikipedia languages have a page describing WikiProjects,\footnote{\url{https://www.wikidata.org/wiki/Q4234303}} and many articles lack infoboxes.\footnote{Only about one-third of English Wikipedia articles have infoboxes per DBPedia's statistics: \url{https://wiki.dbpedia.org/services-resources/ontology}} 
Second, approaches that seek to expand article coverage by also predicting topics for non-annotated articles depend on hand-labeling of topics (which requires language expertise)~\cite{singer2017we,piccardi2020crosslingual,miz2020trending} or language modeling that does not easily scale to all languages on Wikipedia~\cite{asthana2018few}.

In this paper, we make the following contributions:
\begin{itemize}
    \item We present an approach to automatically labeling (almost) all Wikipedia articles across every language of Wikipedia with a consistent set of topics. Specifically, we build on work by Asthana \& Halfaker~\cite{asthana2018few} that uses 64 topics derived from WikiProject tags and extend their language-dependent approach to all Wikipedia languages. The main innovation of our approach is to represent articles in a language-agnostic way using article links that have been mapped to Wikidata items (similar to Piccardi and West~\cite{piccardi2020crosslingual}).
    \item We demonstrate through quantitative and qualitative evaluations that our language-agnostic approach performs equally well or better than alternative approaches.
    \item We release the code and trained model, a dataset of every Wikipedia article and its predicted topics, and APIs for interacting with the models.
\end{itemize}

\section{Related Work}
Three general approaches have been taken to classifying Wikipedia articles into a consistent and coherent set of topics: 1) directly apply existing editor-generated annotations on Wikipedia, 2) linking Wikipedia articles to an external taxonomy, and, 3) learning unsupervised topics and manually labeling them. This work pulls most directly from Section~\ref{rw-annotations} with additional modeling similar to Section~\ref{rw-unsupervised}.

\subsection{Annotations on Wikipedia} \label{rw-annotations}
The most common and simplest strategy for classifying Wikipedia articles by topic is using existing annotations that editors have added to articles. For instance, Wikipedia has a category network that roughly forms a tree with approximately 40 root topics.\footnote{The English Wikipedia system: \url{https://en.wikipedia.org/wiki/Category:Main_topic_classifications}} Not all language communities of Wikipedia, however, use the same set of high-level categories or label articles with categories to the same extent~\cite{lewoniewski2019multilingual}. The category network also is messy, requiring careful rules and pruning to avoid loops when linking a given category to its high-level topic~\cite{piccardi2018structuring,lewoniewski2019multilingual,aghaebrahimian2020testing}.

WikiProjects represent another annotation system where groups of editors interested in a specific topic area---e.g., Medicine~\cite{heilman2011wikipedia}---tag articles that are relevant so that they may be evaluated and improved. These labels have been used directly~\cite{warncke2015misalignment} but also can be aggregated into higher-level topics based on an editor-curated mapping of WikiProjects to topics\footnote{On English Wikipedia: \url{https://en.wikipedia.org/wiki/Wikipedia:WikiProject_Council/Directory}}. Asthana and Halfaker~\cite{asthana2018few} use these aggregated topics as well as building a machine-learning model to predict these topics based on article text for filling in gaps in the annotations and applying them articles that have not yet been labeled. As described in Section~\ref{sec_methods}, this taxonomy of topics and approach of using machine learning to expand existing editor-based annotations is deemed the most appropriate for this task.

\subsection{External Ontologies}
Researchers have also turned to external ontologies---most notably DBPedia~\cite{lehmann2015dbpedia}---for classifying Wikipedia articles. DBPedia has built a manually-curated ontology of around 800 topics that are linked to Wikipedia articles based on which infobox templates are present on a given Wikipedia article. While this has the benefit of being manually-curated and DBPedia also provides linkages to many other ontologies, coverage is limited by how many infobox templates are present and mapped to DBPedia's ontology.

Wikidata offers another ontology that is much more closely-linked to Wikipedia. Wikidata items often either have an instance-of property (P31) or subclass-of property (P279), the network of which can be used to categorize Wikidata items (and their corresponding Wikipedia articles) into a set of high-level topics~\cite{piscopo2018models}. Wikidata's ontology contains loops, dead-ends, and other inconsistencies that limit its usage, however, in applying coherent topics to articles~\cite{brasileiro2016applying,piscopo2018models}. It also is a step removed from Wikipedia articles, which removes the direct connection and feedback loop between the topics applied to an article and what content is included in the article.

\subsection{Unsupervised Approaches} \label{rw-unsupervised}
Some researchers have also avoided these existing ontologies in favor of unsupervised learning of topics and post-hoc labeling. These generally are learned via topic models, most notably Latent Dirichlet Allocation (LDA), with article text as input~\cite{mimno2009polylingual,singer2017we,miz2020trending}. 
These unsupervised approaches have the benefit of generating continuous topic vectors that can be valuable for modeling and having high coverage because they do not rely on annotations. 
However, there are limitations of these approaches for topic labeling of articles systematically. First, the identified latent topics cannot always easily be interpreted in terms of its content. Second, text-based approaches usually require custom adaptations when being applied across all languages due to issues arising from parsing and pre-processing different scripts.

The approach by Piccardi and West~\cite{piccardi2020crosslingual}, which learns a topic model over articles as represented by their links mapped to the language-agnostic Wikidata vocabulary, is very similar to the approach taken in this paper but unsupervised and with additional preprocessing that imputes links that are not there. They also consider the downstream task of supervised topic classification but do so with a balanced dataset, making their performance not comparable to our evaluations with a representative and highly imbalanced dataset.

\section{Methods} \label{sec_methods}
We describe our criteria and chosen approach for developing a language-agnostic approach to labeling Wikipedia articles with a consistent set of topics.

\subsection{Guiding Principles} \label{sec_guidingprinciples}
While high performance is important, it is not the only (or even the most important) criterion that guided the development of the approach described in this paper.

\subsubsection{Coverage}
The main motivation for development of this model was to expand beyond the limited coverage of past approaches in terms of languages and articles within these languages. Knowledge equity, a core component of the Wikimedia Foundation's strategy,\footnote{\url{https://meta.wikimedia.org/wiki/Strategy/Wikimedia_movement/2017/Direction\#Knowledge_equity:_Knowledge_and_communities_that_have_been_left_out_by_structures_of_power_and_privilege}} explicitly calls out the importance of fairly allocating resources, and that includes technologies that can be used to support research and tooling for Wikipedia communities. While this may feel less relevant for e.g., understanding pageview trends in very small wikis (where simply examining the most popular articles may be sufficient), the importance of equity in models such as topic classification is especially salient when considering experiences such as Newcomer Tasks\footnote{\url{https://www.mediawiki.org/wiki/Growth/Personalized_first_day/Newcomer_tasks}} that use topics to provide a personalized, supportive editing experience for new Wikipedia editors. Limiting a model to only languages where pre-trained word embeddings are available or that already have high coverage of annotations would mean that the communities that could benefit the most from growth are cut-off from these technical supports.

To achieve maximal coverage, we use a language-agnostic approach for article \textit{features} (Sec.~\ref{sec:features})
---i.e. the vocabulary for our model is shared across all languages and does not require any language-specific parsing or modeling such as pre-trained word embeddings. Our approach does not consider the actual words in an article at all (just a ``bag-of-links''). As such, it can make topic predictions for (almost) any Wikipedia article, as the only requirement is that it is linked to other Wikipedia articles.

\subsubsection{Size and Simplicity}
We aimed for the model to be fast to train, simple to maintain, low memory requirements, and an overall minimal footprint. These requirements come out of practical concerns---models require maintenance and hardware that must be resourced by the Wikimedia Foundation, which is a non-profit that does not have the money and large staff available to many technology companies. They also come out of a recognition of the environmental costs of training and deploying large machine learning models, a growing trend among technology companies that puts an emphasis on complexity and size instead of seeking simpler, more sustainable solutions~\cite{bender2021dangers}. These concerns are exacerbated by growing evidence that, at least for certain tasks, simpler model architectures combined with appropriate tuning can outperform much more complex models~\cite{joulin2017bag,dacrema2019we,rendle2020neural}. 

This criterion is reflected in our \textit{model architecture} (Sec.~\ref{sec:modelarchitecture}) by choosing a 50-dimensional embeddings (a relatively small dimensionality) and the fastText model architecture, a very simple and low-footprint model architecture~\cite{joulin2017bag}. Our resulting model has only 3,100 parameters (beyond the learned 50-dimensional embeddings), takes approximately 10 minutes to train across 16 CPU threads and takes up only 863MB on disk. Furthermore, the training data for any given article is based solely on the existing links in that article. The model can therefore be trained based on (relatively small) database tables while not processing the much larger and more complex article text dumps.\footnote{\url{https://meta.wikimedia.org/wiki/Data_dumps/FAQ\#How_big_are_the_en_wikipedia_dumps_uncompressed?}} Simpler models also can be more stable by avoiding challenges that tend to arise with larger and more complex models such as underspecification~\cite{d2020underspecification}.

\subsubsection{Feedback}
Given the long history of community-driven decision-making on Wikipedia, including the management of bots and other automated technologies~\cite{geiger2014bots}, machine-learning technologies that are developed to support these editor communities should also be developed with clear pathways for feedback and iteration~\cite{halfaker2020ores} so as to best fulfill the values held by the editor communities they are intended to support~\cite{smith2020keeping}.

We design our model per this criterion in two ways. First, the \textit{features} (Sec.~\ref{sec:features}) in the model are the links in an article. Thus, if the model is not performing as expected for a particular article, the features are easily observable and can be augmented by adding (or removing) any appropriate links in the article. Second, the groundtruth \textit{topic labels} (Sec.~\ref{sec:topiclabels}) used by the model are based on community annotations that are then aggregated to a pre-defined set of topics based on a simple mapping that was derived based on prior efforts by the community. This process is transparent\footnote{The mapping of WikiProject to topic is maintained here: \url{https://github.com/wikimedia/wikitax}} and the labels can be improved through adding appropriate WikiProject tags to articles or adjusting the mapping of WikiProjects to topics. This is in contrast to the category network, which while comprehensive and community-maintained, is very difficult to follow the mapping of category to high-level topic. An alternative approach was also considered that was based not on Wikipedia articles but on classifying their associated Wikidata items as described in Johnson et al.~\cite{johnson2020global}. While this approach is also language-agnostic, simple, efficient, and achieves relatively high coverage, it breaks the clear feedback loop between a particular Wikipedia language edition's article and the topic predictions. It is presented, however, in the quantitative evaluation section for comparison.

\subsubsection{Performance}
The resulting model needs to have high performance so that it is useful for research and supporting editors. The machine-in-the-loop\footnote{See~\cite{clark2018creative} for a delineation of ``machine-in-the-loop'' vs.\ ``human-in-the-loop''.} nature of many Wikipedia tools means that models do not need to be perfect, they just need to be useful for supporting editors in their work. This positioning of the model as support for editors removes the need for an authoritative, state-of-the-art model and instead centers the importance of interpretability and simple feedback loops for adapting and improving the model.

We demonstrate in the \textit{evaluation} (Sec.~\ref{sec:evaluation}) that our model achieves equivalent performance to an already in-use model and thus is acceptable for deployment. We do not introduce greater complexity or footprint in order to extend that performance unnecessarily, though future improvements certainly are always desirable.

\subsection{Our Approach}
Described below are the four components of the chosen model: ground-truth topic labels, input features for representing articles, model architecture, and evaluation technique.

\subsubsection{Topic labels}
\label{sec:topiclabels}
We generate topic labels following the approach by Asthana and Halfaker~\cite{asthana2018few}, which is based on WikiProject tagging. Specifically, WikiProjects, which are groups of editors who organize to focus on specific topic areas,\footnote{e.g., WikiProject Medicine is an exemplar~\cite{heilman2011wikipedia}} tag Wikipedia articles that they deem relevant to their topic area. Almost all articles on English Wikipedia have been tagged by at least one WikiProject and many articles are relevant to multiple WikiProjects. To generate our topic labels, for each article, we identify all its WikiProject tags which can be efficiently extracted via the PageAssessments extension.\footnote{\url{https://www.mediawiki.org/wiki/Extension:PageAssessments\#Database_tables}}
Each of the tags are then mapped to one of 64 topics (such as ``Biography'', ``History and Society'', or ``Mathematics'').\footnote{\url{https://www.mediawiki.org/wiki/ORES/Articletopic\#Taxonomy}} We chose this approach because it was simple, transparent, and Asthana and Halfaker's topic taxonomy was already incorporated into various language-specific tools and research. It has the drawback that it draws its labels solely from English Wikipedia, which we address in the Future Work section. The category network likely represents the best alternative for extending community-driven topics, but lack of hierarchy, inconsistency across languages, and constant evolution pose additional challenges.

For the December 2020 snapshot, this WikiProject-based approach yielded labels for 5,970,598 articles (96\% of all articles on English Wikipedia) and 15,180,479 groundtruth topics (2.5 topics per article). Through the language mappings stored in Wikidata, these labels were also directly associated with articles in any of the approximately 300 other languages on Wikipedia. Thus, the labels for 5,970,598 English Wikipedia articles actually becomes labels for 30,581,076 Wikipedia articles across all languages (55\% of all articles on Wikipedia). A small number of articles were then filtered out if they lacked links that could be mapped to Wikidata IDs and a train/validation/test split of 90\%/2\%/8\% was applied. This lead to 5,088,621 items (27,225,747 articles)\footnote{Item here meaning just English Wikipedia articles while the articles number includes all the articles in other languages about the same concept as well.} in the training data, 113,241 items (603,582 articles) in the validation data, 451,909 items (2,417,700 articles) in the test data, and 25,047,098 articles with data but no labels.

The labels are highly imbalanced ranging from Libraries and Information with 10,043 articles on English Wikipedia to Biographies with 1,862,363 articles on English Wikipedia with the average topic having 235,748 English Wikipedia articles and median topic having 100,240 English Wikipedia articles.

\subsubsection{Features}
\label{sec:features}
We represent each Wikipedia article as an unordered set of links in a language-agnostic way using Wikidata items as the underlying vocabulary.
While the central feature of Wikipedia articles is the actual textual content, an incredibly important and widespread feature of articles is their links. Even stub articles---i.e. those with just a few sentences---often contain several links within the text (or associated templates~\cite{mitrevski2020wikihist}) to other articles that are mentioned or relevant. Almost every article on Wikipedia also is mapped to an associated Wikidata item, which, among other things, stores an editor-curated listing of all the other articles in other Wikipedia language editions that are about the same concept.

Consider the English Wikipedia article for City Bureau,\footnote{\url{https://en.wikipedia.org/wiki/City_Bureau}} a civic journalism non-profit based in Chicago, Illinois, United States. As of January 2021, this article links to 19 other articles in English Wikipedia. These articles, which include other journalistic institutions and Chicago-area places, provide insight into what the City Bureau article itself is about. Furthermore, all of these articles have associated Wikidata items,\footnote{\url{https://en.wikipedia.org/w/api.php?action=query&generator=links&titles=City_Bureau&prop=pageprops&ppprop=wikibase_item&gpllimit=100&gplnamespace=0&redirects&format=json&formatversion=2}} so if there was a City Bureau article in another language that also linked to the same Chicago-area and journalistic concepts, it could be represented in the exact same way.

For every article in every language edition of Wikipedia, we gather all of the links to other articles\footnote{\url{https://www.mediawiki.org/wiki/Manual:Links_table}}. Redirects\footnote{\url{https://www.mediawiki.org/wiki/Manual:Redirect_table}} are resolved, e.g., replacing links to ``Chicago, Illinois'' with the canonical article title ``Chicago''. 
Each article is mapped to its corresponding Wikidata item, \footnote{\url{https://wikitech.wikimedia.org/wiki/Analytics/Data_Lake/Content/Wikidata_item_page_link}}, e.g.,  the article on ``Chicago'' in, both, the English and the German Wikipedia is represented by ``Q1297''\footnote{\url{https://www.wikidata.org/wiki/Q1297}}. 
As a result, the representation of each article as a set of links is language-agnostic.

\begin{figure}[t]
\centering
\includegraphics[width=1\columnwidth]{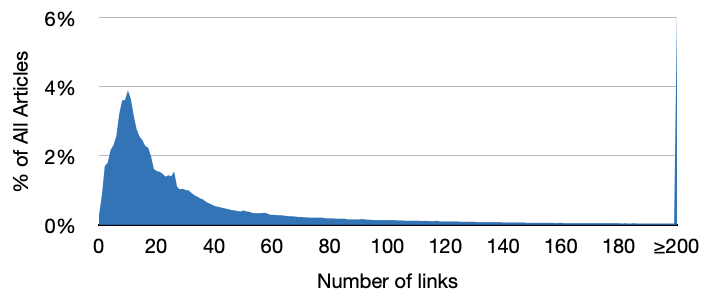}
\Description[Wikipedia Link Distribution]{Bar chart showing what percentage of articles have a given number of links in them. The data ranges from 0 links to greater than or equal to 200 links. There is a steep rise from 0.29\% of articles having zero links to a peak of 3.9\% of articles having 10 links and then drop-off to 1.56\% of articles having 20 links and 1\% of articles having 31 links. The graph slowly declines until the cut-off at 200 links where it then spikes to 6.1\% of articles having 200 or more links.}
\caption{\textbf{Percentage of articles in all of Wikipedia with a given number of links.} Most articles (99.71\%) have at least one link, 77.83\% have 10 or more links. Total links are capped at 200 and 6.10\% of articles have 200 or greater links. Data is based on the 54,248,336 articles across the 306 Wikipedia languages in the December 2020 dumps. Articles include all non-redirects in namespace 0. Links are only counted if they go to another article in namespace 0 in that wiki.}
\label{fig:link_distribution}
\end{figure}

For the December 2020 snapshot, this yielded 55,334,958 articles that aggregated to 20,119,210 distinct concepts (Wikidata items). These articles were associated with 3,122,037,115 links that were mapped to a vocabulary of 18,037,708 distinct Wikidata items (though only 4,145,064 of these items were retained in the final model's vocabulary). As described in the next section, about half of these articles have associated groundtruth labels and can therefore be used for training and testing. The distribution of links per article is shown in Figure~\ref{fig:link_distribution}. As can be seen, the vast majority of articles (99.71\%) have at least one link and thus can be covered by the model.

\subsubsection{Model Architecture}
\label{sec:modelarchitecture}
We model the problem of topic classification as a multi-class supervised learning task.
We specifically chose the fastText supervised classification architecture~\cite{joulin2017bag}, which has been demonstrated to be efficient and highly effective at (multi-class) classification tasks. The bag-of-links approach we take here is functionally equivalent to the bag-of-words approach the fastText model was designed to use as input.

The fastText model learns embeddings for each word in its vocabulary, averages the embeddings together for all the words in a document, and then trains a simple, one-vs-all logistic regression over this average embedding to make its predictions. It requires the following hyperparameters: embedding dimensionality, threshold for how many times a link must appear to be included in the model vocabulary, learning rate, and number of epochs. We fix the learning rate at $0.1$ and conduct grid search over dimensionality (${50, 100}$), vocabulary threshold (${5, 10, 20}$), and epochs (${2,3}$).\footnote{The number of epochs may seem low but remember that an article may exist in many languages, so the model practically speaking sees these training examples more than once per epoch.} We saw minimal variation (\textasciitilde1\%) in F1 scores and so use the lowest-footprint values: dimensionality of $50$, epochs of $2$, and vocabulary threshold of $20$. The model also has a window-size hyperparameter used for training, but given that there is no order to the bag-of-links we use as input features,\footnote{Links do have an order in an article, but our approach is unaware of this order and would require an additional and much lengthier preprocessing step to extract that information from an article's text.} we randomize the link order in the training data and set the window size to be relatively large ($20$) to reduce the chance of learning any spurious signals.

\subsubsection{Evaluation}
\label{sec:evaluation}
We evaluate the resulting model in two different ways: a quantitative evaluation using held-out test data and qualitative evaluation of predictions by the model in several languages by expert Wikipedians. Details about how to access the code, data, and model for further evaluation or usage are included in Appendix as well.

For the quantitative evaluation, we use a standard held-out test set of data. Given that articles can appear in multiple languages, we randomly sort articles into the train-validation-test splits based on their associated Wikidata item---i.e. if the English article for Curt Flood\footnote{\url{https://en.wikipedia.org/wiki/Curt_Flood}} is randomly placed into the test set, then any other language versions of that article\footnote{In this case, French, Arabic, Egyptian Arabic, Japanese, and Polish as can be seen under sitelinks on Wikidata: \url{https://www.wikidata.org/wiki/Q5195372}} will also be placed in the test set. Even though the other language versions of the article will very likely have different links associated with them~\cite{hecht2010tower}, they share labels and we choose to not consider them to be independent data points. Given the heavy imbalance in data, we report both micro and macro precision, recall, and F1 statistics. We also report average precision, which summarizes the precision-recall curve and is therefore threshold-agnostic (the other approaches uses the default threshold of 0.5 for mapping probabilities to binary predictions).

The results for our language-agnostic link-based model are also compared to two other models. The first is the Wikidata-based model described in Johnson et al.~\cite{johnson2020global}, which predicts topics for a Wikipedia article based on the properties and values contained in its associated Wikidata item. It is trained with the same fastText architecture as the language-agnostic links-based model, a learning rate of $0.1$, $25$ epochs, $50$ dimensions, and a vocabulary threshold of $3$. The second comparison model is the language-dependent, text-based model described in Asthana and Halfaker~\cite{asthana2018few}, which predicts topics for a Wikipedia article based on a gradient-boosted classifier learned over top a document embedding formed by averaging together word embeddings associated with each word in the article. This is very similar to the fastText architecture and an API for the model is available through the Wikimedia Foundation.\footnote{\url{https://www.mediawiki.org/wiki/ORES\#Topic_routing}}

The different training paradigms and choice of input features make it difficult to directly compare these different approaches under the exact same conditions.\footnote{While all three models share the same set of labels and groundtruth data, the choice of input features---article text, article links, Wikidata statements---means that each model has a different number of possible training examples.} We instead choose to compare these models as they would be deployed so as to directly compare their potential value to the Wikimedia community.

For the qualitative evaluation, we focus on five languages chosen for their geographic coverage and the willingness of their communities to experiment with new tools: Arabic (ar), Czech (cs), English (en), French (fr), and Vietnamese (vi). For each language, we generate ten random examples of articles that are confidently\footnote{We use the threshold of 0.5 though its likely that better results would be achieved with topic-specific thresholds} labeled with each topic. These examples are presented to Wikipedians who are familiar with that language and they evaluate whether the topic is relevant to the article or not (a binary yes/no with the opportunity for explanation). Across the 64 topics, 5 languages, and 10 articles per topic, this results in 3,200 evaluations. The same evaluation is done (by the same Wikipedians) for the language-dependent, text-based model in each language. The Wikidata-based model was left out of this evaluation because it did not clearly outperform the language-agnostic link-based model in the quantitative evaluation.

\section{Results}
We first look at the quantitative results for just English-language articles for all three models as the most-directly comparable set of results (Rows 1-3 of Table~\ref{tab:en_results}). The models perform very similarly (within 2-3\% of each other) though the text-based model performs the best in English and the Wikidata model has lower recall (and therefore F1 and average precision). Looking at the fourth row in Table~\ref{tab:en_results}, we see that the language-agnostic link-based model does not suffer in performance when making predictions for articles in other languages. Together, these results provide evidence that this model is able to provide both high performance and high coverage.

\begin{table*}[t]
\centering
\small
\begin{tabular}{l|cccccccc}
Model & Pre. (micro) & Pre. (macro) & Rec. (micro) & Rec. (macro) & F1 (micro) & F1 (macro) & Avg. Pre. (micro) & Avg. Pre. (macro) \\
\hline
Wikidata & 0.891 & 0.854 & 0.796 & 0.679 & 0.838 & 0.752 & 0.895 & 0.797 \\
Link-Based (en) & 0.866 & 0.826 & 0.808 & 0.719 & 0.835 & 0.766 & 0.892 & 0.814 \\
Text-Based (en) & 0.883 & 0.846 & 0.817 & 0.723 & 0.847 & 0.776 & 0.908 & 0.841 \\
\hline
Link-Based (all) & 0.877 & 0.836 & 0.793 & 0.678 & 0.83 & 0.744 & 0.891 & 0.795 \\
\end{tabular}
\Description[Summary of model results]{A table that shows the performance of each of the models (Wikidata, Link-based, and text-based) on English Wikipedia. They perform similarly and a fourth row shows that the link-based model shows similar performance across all of Wikipedia.}
\caption{\textbf{Overall model results.} Model evaluation results for the three models. The first three rows show results just for test data from English Wikipedia. The two entries for the link-based language-agnostic model are for the same model but different test sets: held-out English-language articles only (row 2) and held-out articles from all languages (row 4). Average-precision summarizes the precision-recall curve for all possible thresholds at which a prediction probability may be turned into a binary yes. See \url{https://scikit-learn.org/stable/modules/generated/sklearn.metrics.average_precision_score.html} for more information.}
\label{tab:en_results}
\end{table*}

The quantitative results are limited, however, by the fact that articles in languages outside of English only will have groundtruth labels if they have an English-language equivalent, which is not true for almost half of the articles. The results from the qualitative evaluation of English and four additional languages (Arabic, Czech, French, Vietnamese) provide some insight into whether the model generally seems to work for non-English languages.

We see that the link-based model consistently outperforms the language-specific, text-based model in all five languages tested (Table~\ref{tab:qual_summary}). The increase in performance varies by language but ranges from 4\% (absolute) better in Arabic and English to almost 12\% better in Vietnamese. There are certain topics where the language-specific, text-based model performs better (though the low sample size per topic makes it difficult to read too far into this) such as the women biographies topic. This suggests that for certain topics, especially given the sensitive nature of gender, relying on Wikidata's information may be a far better solution. Finally, the feedback from raters highlighted the limitations of some of these topics---e.g., in one instance, while all of the articles in the topic for Northern Africa were correct, they all came from the same country (Morocco), which may reflect bias in the wiki or bias in the recall of the model. Either way, some topics are clearly too high-level for certain use-cases and complementary models will be required to address these. The results also serve to verify the precision values seen in the quantitative evaluations.

\begin{table}[t!]
\centering
\begin{tabular}{l|cc}
Model & Overall Precision & Topics < 7/10 \\
\hline
Arabic (LD) & 90.9\% & 4 \\
Arabic (LA) & 94.7\% & 2 \\
\hline
Czech (LD) & 74.5\% & 17 \\
Czech (LA) & 81.4\% & 11 \\
\hline
English (LD) & 85.5\% & 7 \\
English (LA) & 89.5\% & 6 \\
\hline
French (LD) & 79.7\% & 11 \\
French (LA) & 88.0\% & 5 \\
\hline
Vietnamese (LD) & 79.7\% & 14 \\
Vietnamese (LA) & 91.3\% & 2 \\
\end{tabular}
\Description[Comparison table for language-dependent and language-agnostic models]{A comparison of the overall precision for each model type shows that across Arabic, Czech, English, French, and Vietnamese, the language-dependent model performs worse than the language-agnostic model both in terms of overall precision and the number of topics where precision was lower than 70\%.}
\caption{\textbf{Evaluation of quality of predictions by topic for language-agnostic and language-dependent models.} This table summarizes the qualitative evaluation of the language-agnostic (LA) model and language-dependent (LD) model. It reports the overall precision across the 64 topics (10 articles per topic) for each model as well as how many of the 64 topics did not have at least 70\% precision (7 out of 10 correct). The language-agnostic model outperforms the language-dependent model for all languages in terms of both metrics. Keep in mind that while the language-agnostic model is a single model, there are in fact five language-dependent models, one for each language, with their own word embeddings and model parameters.}
\label{tab:qual_summary}
\end{table}

\section{Discussion}
We developed a language-agnostic topic classification model that can automatically label (almost) all Wikipedia articles across all languages. The results demonstrate that the advantages of this approach by meeting the criteria laid out in Section~\ref{sec_guidingprinciples} around coverage, simplicity, and feedback, while at the same time performing with high precision and recall. In fact, the evaluation showed that the model performs equally well or better than alternative approaches. 

More generally, we believe language-agnostic approaches to building machine-learning models in the service of Wikimedia projects can be applied to many other classification tasks. For instance, labeling the quality of articles is a common maintenance task on Wikipedia but one that is difficult to constantly keep up-to-date~\cite{warncke2013tell}. While not all language editions use the same set of quality classes~\cite{lewoniewski2019multilingual}, Lewoniewski et al.~\cite{lewoniewski2017relative} built a simple language-agnostic quality model that can be applied to any language. While certain tasks such as vandalism detection may prove too nuanced for language-agnostic models, future work should explore how to apply language-agnostic methods to other tasks such as identifying more specific topics like the countries associated with an article or the intention behind edits~\cite{yang2017identifying}.

\section{Future Work}
While we demonstrate that language-agnostic topic classification model presented here is performant and adheres to the principles laid out for its development, there is still plenty of room for improvement. The clearest downside to the approach presented is its reliance on groundtruth labels and a taxonomy derived solely from English Wikipedia. A few other language editions (Arabic, Hungarian, Turkish, French) use the PageAssessments extension that allows for easy extraction of WikiProject tags and many other languages have active WikiProjects whose data could be mined through other methods. This would directly provide some labels for articles that do not exist in English Wikipedia but have been tagged by WikiProjects that have corollaries on English Wikipedia (as determined by Wikidata-maintained sitelinks). Even better would be working with these communities to map their other WikiProjects to the topic taxonomy and propose adjustments to the taxonomy, which may itself reflect English-language and cultural biases. Given the evaluations presented here, we think this taxonomy has utility in languages outside of English, but it is open to iteration.

While almost all articles have links, many have only a few, making it difficult for the model to predict topics. There are language-agnostic approaches to inferring what links might be missing from a given Wikipedia article---e.g., the approach presented by Piccardi and West~\cite{piccardi2020crosslingual}---but more work is needed to ensure that that process is efficient. Their work applied it across 28 languages and still required language-specific parsing of the article text, but it is a promising approach. In the meantime, tools such as link recommendation for new editors\footnote{\url{https://www.mediawiki.org/wiki/Growth/Personalized_first_day/Structured_tasks/Add_a_link}} can be used to help add links directly to articles where they are most needed.

We considered only one model architecture for a language-agnostic model in this paper. There are many other potential language-agnostic architectures though such as graph-based approaches~\cite{moskalenko2020scalable} or applying sequence models to reader navigation~\cite{sen2017cartograph}. While not all of these language-agnostic models are simple and adhere closely to the guiding principles for this model, these other approaches may complement the simple bag-of-links approach described here and be necessary for achieving sufficient performance and coverage on other tasks.

\begin{acks}
We would like to thank the many supporters of this language-agnostic topic classification project and the Wikipedians who have built out such extensive annotation systems on Wikipedia. In particular, Aaron Halfaker for his constant support and guidance in building on his initial work and Marshall Miller, Habib Mhenni, Beno\^it Evellin, Martin Urbanec, Bluetpp, and HAKSOAT for their support in evaluating these models.
\end{acks}

\bibliographystyle{ACM-Reference-Format}
\bibliography{bibliography}

\appendix
\section{Availability}
We make as much of the code, data, and predictions open for further usage and critique. All of the topics predicted for every single Wikipedia article are available for download and use.\footnote{\url{https://figshare.com/articles/dataset/Wikipedia_Article_Topics_for_All_Languages_based_on_article_outlinks_/12619766}} The code for the model is available under open license,\footnote{\url{https://github.com/geohci/wikipedia-language-agnostic-topic-classification}} though the data pipeline is written for a Hadoop cluster operated by the Wikimedia Foundation and thus likely would require adjustments to use the public dump files. For interested users, an API to compare the different models is made available via a simple interface.\footnote{\url{https://wiki-topic.toolforge.org/comparison}}

\section{Complete Results}

\begin{table*}[th]
\centering
\footnotesize
\begin{tabular}{l|ccccccccc}
Topic & n & TP & FP & TN & FN & Precision & Recall & F1 & Average Precision \\
\hline
europe & 783307 & 681220 & 90139 & 1544254 & 102087 & 0.883 & 0.870 & 0.876 & 0.952 \\
biography & 677093 & 621664 & 49035 & 1691572 & 55429 & 0.927 & 0.918 & 0.922 & 0.971 \\
stem & 480236 & 431709 & 24123 & 1913341 & 48527 & 0.947 & 0.899 & 0.922 & 0.973 \\
asia & 360865 & 304097 & 35810 & 2021025 & 56768 & 0.895 & 0.843 & 0.868 & 0.933 \\
sports & 334774 & 315905 & 10497 & 2072429 & 18869 & 0.968 & 0.944 & 0.956 & 0.980 \\
media & 327692 & 281754 & 37092 & 2052916 & 45938 & 0.884 & 0.860 & 0.872 & 0.938 \\
western-europe & 290047 & 243700 & 28115 & 2099538 & 46347 & 0.897 & 0.840 & 0.867 & 0.944 \\
north-america & 282981 & 216595 & 40953 & 2093766 & 66386 & 0.841 & 0.765 & 0.801 & 0.895 \\
biology & 247481 & 232241 & 6260 & 2163959 & 15240 & 0.974 & 0.938 & 0.956 & 0.985 \\
geographical & 213947 & 145537 & 33160 & 2170593 & 68410 & 0.814 & 0.680 & 0.741 & 0.825 \\
eastern-europe & 176961 & 149351 & 14603 & 2226136 & 27610 & 0.911 & 0.844 & 0.876 & 0.938 \\
southern-europe & 172638 & 138080 & 19706 & 2225356 & 34558 & 0.875 & 0.800 & 0.836 & 0.913 \\
northern-europe & 161724 & 118894 & 21354 & 2234622 & 42830 & 0.848 & 0.735 & 0.787 & 0.869 \\
history & 142121 & 83506 & 20703 & 2254876 & 58615 & 0.801 & 0.588 & 0.678 & 0.757 \\
music & 127211 & 110574 & 9764 & 2280725 & 16637 & 0.919 & 0.869 & 0.893 & 0.939 \\
women & 113835 & 47597 & 20674 & 2283191 & 66238 & 0.697 & 0.418 & 0.523 & 0.612 \\
films & 112002 & 90799 & 15972 & 2289726 & 21203 & 0.850 & 0.811 & 0.830 & 0.904 \\
east-asia & 97601 & 80108 & 10181 & 2309918 & 17493 & 0.887 & 0.821 & 0.853 & 0.909 \\
military-and-warfare & 103937 & 64034 & 15022 & 2298741 & 39903 & 0.810 & 0.616 & 0.700 & 0.774 \\
politics-and-government & 95746 & 53169 & 15573 & 2306381 & 42577 & 0.773 & 0.555 & 0.646 & 0.724 \\
west-asia & 91404 & 73965 & 8718 & 2317578 & 17439 & 0.895 & 0.809 & 0.850 & 0.914 \\
philosophy-and-religion & 89395 & 51774 & 14468 & 2313837 & 37621 & 0.782 & 0.579 & 0.665 & 0.717 \\
visual-arts & 86498 & 51830 & 14574 & 2316628 & 34668 & 0.781 & 0.599 & 0.678 & 0.744 \\
transportation & 84279 & 71175 & 6045 & 2327376 & 13104 & 0.922 & 0.845 & 0.881 & 0.921 \\
literature & 75739 & 41881 & 11840 & 2330121 & 33858 & 0.780 & 0.553 & 0.647 & 0.718 \\
south-asia & 72071 & 60734 & 5569 & 2340060 & 11337 & 0.916 & 0.843 & 0.878 & 0.919 \\
africa & 71004 & 49691 & 8820 & 2337876 & 21313 & 0.849 & 0.700 & 0.767 & 0.835 \\
south-america & 61705 & 46888 & 7646 & 2348349 & 14817 & 0.860 & 0.760 & 0.807 & 0.877 \\
north-asia & 63870 & 47955 & 7789 & 2346041 & 15915 & 0.860 & 0.751 & 0.802 & 0.871 \\
oceania & 60421 & 46368 & 5013 & 2352266 & 14053 & 0.902 & 0.767 & 0.829 & 0.876 \\
business-and-economics & 53360 & 25192 & 9361 & 2354979 & 28168 & 0.729 & 0.472 & 0.573 & 0.614 \\
technology & 44378 & 25983 & 8303 & 2365019 & 18395 & 0.758 & 0.585 & 0.661 & 0.721 \\
engineering & 45147 & 30979 & 4465 & 2368088 & 14168 & 0.874 & 0.686 & 0.769 & 0.820 \\
architecture & 42920 & 22882 & 7432 & 2367348 & 20038 & 0.755 & 0.533 & 0.625 & 0.684 \\
medicine-and-health & 41775 & 28278 & 4943 & 2370982 & 13497 & 0.851 & 0.677 & 0.754 & 0.815 \\
earth-and-environment & 40405 & 26281 & 4867 & 2372428 & 14124 & 0.844 & 0.650 & 0.735 & 0.779 \\
television & 38695 & 26503 & 5099 & 2373906 & 12192 & 0.839 & 0.685 & 0.754 & 0.806 \\
society & 38816 & 10600 & 6410 & 2372474 & 28216 & 0.623 & 0.273 & 0.380 & 0.407 \\
southeast-asia & 38135 & 27982 & 3772 & 2375793 & 10153 & 0.881 & 0.734 & 0.801 & 0.856 \\
space & 36009 & 32368 & 1331 & 2380360 & 3641 & 0.961 & 0.899 & 0.929 & 0.960 \\
linguistics & 32034 & 20520 & 2452 & 2383214 & 11514 & 0.893 & 0.641 & 0.746 & 0.772 \\
computing & 25619 & 18487 & 3769 & 2388312 & 7132 & 0.831 & 0.722 & 0.772 & 0.837 \\
central-america & 24075 & 14708 & 2659 & 2390966 & 9367 & 0.847 & 0.611 & 0.710 & 0.762 \\
entertainment & 22751 & 9633 & 4024 & 2390925 & 13118 & 0.705 & 0.423 & 0.529 & 0.587 \\
internet-culture & 23239 & 17083 & 2064 & 2392397 & 6156 & 0.892 & 0.735 & 0.806 & 0.872 \\
education & 23323 & 5774 & 3312 & 2391065 & 17549 & 0.635 & 0.248 & 0.356 & 0.381 \\
chemistry & 22115 & 16800 & 2614 & 2392971 & 5315 & 0.865 & 0.760 & 0.809 & 0.884 \\
northern-africa & 20544 & 11847 & 3437 & 2393719 & 8697 & 0.775 & 0.577 & 0.661 & 0.706 \\
food-and-drink & 19547 & 11808 & 2595 & 2395558 & 7739 & 0.820 & 0.604 & 0.696 & 0.731 \\
performing-arts & 17512 & 8030 & 2796 & 2397392 & 9482 & 0.742 & 0.459 & 0.567 & 0.587 \\
physics & 16666 & 9738 & 2722 & 2398312 & 6928 & 0.782 & 0.584 & 0.669 & 0.730 \\
books & 17010 & 9479 & 2561 & 2398129 & 7531 & 0.787 & 0.557 & 0.653 & 0.686 \\
video-games & 16301 & 14382 & 700 & 2400699 & 1919 & 0.954 & 0.882 & 0.917 & 0.947 \\
mathematics & 15628 & 10965 & 2110 & 2399962 & 4663 & 0.839 & 0.702 & 0.764 & 0.820 \\
eastern-africa & 16386 & 10657 & 1831 & 2399483 & 5729 & 0.853 & 0.650 & 0.738 & 0.793 \\
comics-and-anime & 14774 & 11416 & 1248 & 2401678 & 3358 & 0.901 & 0.773 & 0.832 & 0.862 \\
software & 14288 & 8377 & 3391 & 2400021 & 5911 & 0.712 & 0.586 & 0.643 & 0.692 \\
western-africa & 14481 & 9804 & 1640 & 2401579 & 4677 & 0.857 & 0.677 & 0.756 & 0.816 \\
southern-africa & 10012 & 6318 & 1018 & 2406670 & 3694 & 0.861 & 0.631 & 0.728 & 0.758 \\
central-asia & 9549 & 5304 & 1602 & 2406549 & 4245 & 0.768 & 0.555 & 0.645 & 0.687 \\
central-africa & 6698 & 3932 & 985 & 2410017 & 2766 & 0.800 & 0.587 & 0.677 & 0.722 \\
fashion & 5789 & 2679 & 892 & 2411019 & 3110 & 0.750 & 0.463 & 0.572 & 0.579 \\
radio & 4373 & 2349 & 522 & 2412805 & 2024 & 0.818 & 0.537 & 0.649 & 0.636 \\
libraries-and-information & 3765 & 1449 & 537 & 2413398 & 2316 & 0.730 & 0.385 & 0.504 & 0.470 \\
\end{tabular}
\Description[Topic-specific results for the language-agnostic model]{A large table that reports the sample size, confusion matrix, precision, recall, F1, and average precision for each topic in the test set. Most topics perform well with the exception of Education.}
\caption{\textbf{Topic-specific results for the language-agnostic model.} For each topic, the number of test articles (n), true positives (TP), false positives (FP), true negatives (TN), false negatives (FN), precision, recall, F1, and average precision (Ave.\ Pre.) are given. Average-precision summarizes the precision-recall curve for all possible thresholds at which a prediction probability may be turned into a binary yes. See \url{https://scikit-learn.org/stable/modules/generated/sklearn.metrics.average_precision_score.html} for more information. The model performs quite well for almost all topics (precision over 0.7) with the exception of Education. Recall suffers for several topics as well, mostly those with fewer examples.}
\label{tab:topic_eval_results}
\end{table*}

\end{document}